\documentclass[prd,nofootinbib,floats,aps,twocolumn,tightenlines,superscriptaddress]{revtex4-2}
\usepackage{natbib}
\usepackage{graphicx}
\usepackage{mathtools}
\usepackage[normalem]{ulem}
\usepackage{hyperref}
\hypersetup{
	colorlinks=true,
	linkcolor=blue,
	filecolor=magenta,      
	urlcolor=cyan,
}
\usepackage{commath}
\usepackage{bm}
\usepackage{amsfonts,amsmath,amssymb,amsthm}
\usepackage[usenames,dvipsnames]{xcolor}
\usepackage{dsfont}
\usepackage{enumitem}
\usepackage{color}
\usepackage{tikz}
\usepackage{physics}
\usepackage{verbatim}
\DeclareMathOperator{\sgn}{sgn}
\newcommand{\ii}{\mathrm{i}}

\renewcommand*\d[2][]{%
	\mathrm{d}%
	\ifx\relax#1\relax\else
	\rule{-0.02em}{1.5ex}^{#1}\rule{0.08em}{0ex}\!
	\fi
	#2\,
}

\newcommand{\tcr}{\textcolor{red}}
\newcommand{\tcb}{\textcolor{blue}}

\makeatletter 

\renewcommand\onecolumngrid{
\do@columngrid{one}{\@ne}%
\def\set@footnotewidth{\onecolumngrid}
\def\footnoterule{\kern-6pt\hrule width 1.5in\kern6pt}%
}

\renewcommand\twocolumngrid{
        \def\footnoterule{
        \dimen@\skip\footins\divide\dimen@\thr@@
        \kern-\dimen@\hrule width.5in\kern\dimen@}
        \do@columngrid{mlt}{\tw@}
}%
\makeatother

\newcommand\restr[2]{{
		\left.\kern-\nulldelimiterspace 
		#1 
		\vphantom{\normal|} 
		\right|_{#2} 
}}

\newcommand{\diff}{\mathrm{d}}

\newcommand{\R}[1]{\mathbb{R}^{#1}}

\newcommand{\beq}{\begin{equation}}
	\newcommand{\eeq}{\end{equation}}

\newtheorem*{definition*}{Definition}

\newtheorem*{claim*}{Claim}
\newtheorem{criterion}{Criterion}

\begin{document}
	
\title{A proposal to characterize and quantify superoscillations}

\author{Yu Li}
\email{yu.li@uwaterloo.ca}
\affiliation{Department of Applied Mathematics, University of Waterloo, Waterloo, Ontario, N2L 3G1, Canada}
\affiliation{Institute for Quantum Computing, University of Waterloo, Waterloo, Ontario, N2L 3G1, Canada}
	
\author{Jos\'{e} Polo-G\'{o}mez}
\email{jpologomez@uwaterloo.ca}
\affiliation{Department of Applied Mathematics, University of Waterloo, Waterloo, Ontario, N2L 3G1, Canada}
\affiliation{Institute for Quantum Computing, University of Waterloo, Waterloo, Ontario, N2L 3G1, Canada}
\affiliation{Perimeter Institute for Theoretical Physics, Waterloo, Ontario, N2L 2Y5, Canada}
	
\author{Eduardo Mart\'{i}n-Mart\'{i}nez}
\email{emartinmartinez@uwaterloo.ca}
\affiliation{Department of Applied Mathematics, University of Waterloo, Waterloo, Ontario, N2L 3G1, Canada}
\affiliation{Institute for Quantum Computing, University of Waterloo, Waterloo, Ontario, N2L 3G1, Canada}
\affiliation{Perimeter Institute for Theoretical Physics, Waterloo, Ontario, N2L 2Y5, Canada}

\begin{abstract}
 
We present a formal definition of superoscillating function. We discuss the limitations of previously proposed definitions and illustrate that they do not cover the full gamut of superoscillatory behaviours. We demonstrate the suitability of the new proposal with several examples of well-known superoscillating functions that were not encompassed by previous definitions.
		
\end{abstract}
	
\maketitle
	
\section{Introduction}\label{Section: introduction}
	
The concept of superoscillations originated in~\cite{Aharonov1991} (although it was anticipated in~\cite{Wolter1950,Wolter1950Translation,DiFrancia1952,Nye1974}), and its mathematical foundations were first laid out in~\cite{Berry1995}. A great part of their appeal resides on their paradoxical nature: a bandlimited function (i.e., a function with a bounded Fourier spectrum) can, in a bounded interval, oscillate faster than its fastest Fourier component. Even more strikingly, this ``superoscillatory behaviour'' can be arbitrarily fast, and the intervals over which this phenomenon happens can be arbitrarily long (see, e.g.,~\cite{Berry2019Roadmap}). 

Besides the perplexity that the notion induces on those who nowadays confront it for the first time, and despite the relative youth of the field, superoscillations have already had an impact across multiple areas of physics, ranging from the foundations of quantum mechanics~\cite{Aharonov1988,Aharonov1991, Aharonov1993,Berry1994Billiards,Aharonov2002tunneling,Aharonov2003tunnelingweak,Berry2017,Aharonov2020Conservationlaws,Aharonov2023Conservationlaws2,Berry2010,Berry2011Spin,Berry2011Pointer,Yuan2016,Kempf2004,Kempf2017} to classical and quantum information~\cite{Kempf2000Beethoven,Reznik2005,Ferreira2006,Ferreira2007,Hao2008,Pye2015,Ber2015,Tang2016,Kempf2018Fouraspects}. They have also sparked remarkable technical applications, such as a way of working around the diffraction limit to improve optical imaging and focusing~\cite{Huang2008,Zheludev2008,Huang2009,Berry2013Nonparaxial,Wang2015,Qin2016,Yuan2017,Li2018,Luo2018,Yuan2019,Baumgartl2011,Lindberg2012,Rogers2013Imaging,Rogers2013Needle,Dong2017,Rogers2020,Zheludev2021}, and developing ultrafast optics~\cite{Eliezer2014,Manzoni2015,Eliezer2017,Eliezer2018}, among many others. 

The properties of superoscillations have been the object of extensive study, both from a mathematical~\cite{Berry2008,Dennis2009,Aharonov2011Math,Aharonov2012SO3,Chojnacki2016,Aoki2018,Colombo2018,Aharonov2018Classes,Colombo2019,Aharonov2021,Colombo2023Generating,Colombo2023General} and a physical~\cite{Ferreira2002,Calder2004,Berry2006,Dennis2008,Berry2013,Katzav2013,Aharonov2013,Lee2014,Buniy2014,Aharonov2015GenSchrodinger,Berry2016,Soda2020,Aharonov2020KG,Aharonov2022Schrodinger} perspective. However, even with all the progress that has been made so far, a clear-cut, rigorous definition of what it means for a function to display superoscillations is still arguably missing. One research line that has approached the problem of studying the mathematical properties of superoscillations (in a systematic, mathematically rigorous way) resorts to defining the concept of superoscillating sequences (see, e.g.,~\cite{Aharonov2011Math,Aharonov2017Monograph}), rather than defining what it means for a given, single function to display superoscillations. However, we will argue that this way of approaching superoscillations can be too restrictive. Most of the other lines of study, as far as we are aware of, have traditionally either used the most compelling and graphically self-evident examples as guiding principles (such as~\cite{Katzav2013,Chojnacki2016,Berry2017}), or used a notion of `local wavenumber' that leads to a definition~\cite{Berry2008,Dennis2008,Dennis2009} that suffers from limitations and ambiguities, and does not identify superoscillations in some simple examples, as we will see in this paper. 

Here, we propose a formal definition of what it means for a real-valued bandlimited function to display superoscillatory behaviour in a particular interval. This definition can be applied unambiguously to decide if any given function is superoscillating in any specified interval. We will also argue that the definition we introduce captures the physically relevant notion of superoscillations.

The paper is organized as follows: in Sec.~\ref{Section: The challenge of defining superoscillations}, we review the two main descriptions of superoscillations that have been previously used in the literature, and discuss the problems they exhibit. In Sec.~\ref{Section: Identifying 1D superoscillations}, we present two criteria to identify superoscillations, and we then use them to formulate a rigorous definition of superoscillating function in an interval. Sec.~\ref{Section: Examples} is dedicated to illustrating how the proposed definition identifies and quantifies superoscillations by testing it against several examples. Finally, we present our conclusions in Sec.~\ref{Section: Conclusion}. 

\vspace{-0.15cm}

\section{The challenge of defining superoscillations}\label{Section: The challenge of defining superoscillations}

\vspace{-0.15cm}

\begin{figure}
\includegraphics[scale=0.9]{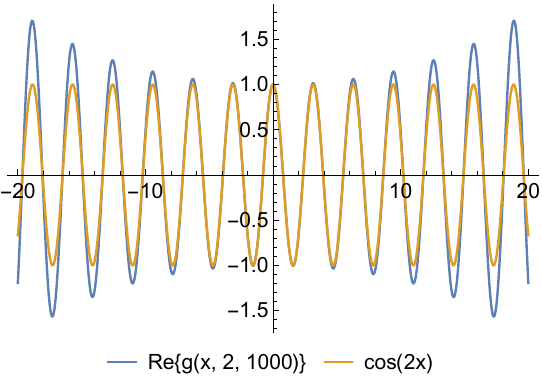}
\caption{Comparison between the real part of $g(x, 2, 1000)$ and $\cos(2x)$, showing superoscillations around $x=0$.}
\label{Fig: Classic superoscillation example}
\end{figure}

Perhaps surprisingly, after over 30 years of study, there is still no consensus on a rigorous definition of superoscillating function. In fact, the concept of superoscillations is usually introduced in the literature in a predominantly qualitative way. The reason for this is probably that it is actually very easy to provide an intuitive notion of superoscillation: a superoscillating function is a bandlimited function (i.e., with bounded Fourier transform) that exhibits a region (the superoscillating interval or set) where it oscillates with a frequency outside of its bandwidth. More colloquially, we would say that there are intervals in the graph of the function where it oscillates ``faster than its fastest Fourier component''~\cite{Berry2019Roadmap}. This intuitive notion is often complemented with the most preeminent example of superoscillating function~\cite{Aharonov1991,Berry2006,Berry2019Roadmap}:
\begin{align}
&g(x,a,N) \coloneqq \bigg[ \cos\bigg(\frac{x}{N}\bigg) + \ii a \sin\bigg(\frac{x}{N}\bigg) \bigg]^N \label{Eq: definition classic example}\\
& = \sum_{j=0}^{N} \frac{1}{2^N} \binom{N}{j}(1+a)^{N-j} (1-a)^{j}  e^{\ii(1-2 j/N) x}. \label{Eq: series classic example}
\end{align}
As can be seen explicitly from Eq.~\eqref{Eq: series classic example}, $g$ is bandlimited to the interval $[-1,1]$. However, from Eq.~\eqref{Eq: definition classic example} it is also easy to see that near the origin \mbox{$g(x,a,N) \approx e^{\ii a x}$}. Thus, for $|a|>1$, $f$ will exhibit superoscillations in a neighbourhood of $x=0$, which is usually estimated to be the region $|x|<\sqrt{N}$ (see, e.g.,~\cite{Berry2006,Berry2019Roadmap}). The superoscillatory behaviour of this family of functions is illustrated in Fig.~\ref{Fig: Classic superoscillation example}, where $\Re\{g(x,2,1000)\}$ and $\cos(2x)$ are shown to oscillate synchronically around $x = 0$.  

One particular feature of the family of examples given by $g(x,a,N)$ is the fact that
\begin{equation}
\lim_{N \to \infty} g(x,a,N) = e^{\ii a x}, \quad \forall \,x\in\R{}.
\end{equation}
This convergence is, of course, pointwise in the whole real line, but it is also uniform when we restrict ourselves to a fixed compact interval. This property motivated the introduction of the concept of superoscillating sequence, which paved the way for a significant amount of the past and ongoing investigation of superoscillations from the mathematical point of view (see, e.g.,~\cite{Aharonov2011Math,Aharonov2017Monograph} for reviews): following~\cite{Aharonov2017Monograph}, a generalized Fourier sequence $\{Y_n\}$ is a sequence of functions of the form
\begin{equation}
Y_n(x,a) = \sum_{j=0}^n C_j(n,a)\,e^{\ii k_j(n) x},
\end{equation}
where $n \in \mathbb{N}$, $x \in \R{}$, $a \in \R{^+}$, and $C_j$ and $k_j$ are real-valued functions, respectively. Such a sequence is then considered a \textit{superoscillating sequence} in a compact set $K \subset \R{}$ if 
\begin{equation}\label{Eq: condition superoscillating sequence}
\lim_{n\to\infty}Y_n(x,a) = e^{\ii g(a) x}
\end{equation}
uniformly in $K$, with $|g(a)| > \sup_{j,n}|k_j(n)|$. 

This definition, while formal, is still not a definition of superoscillating function. It defines specific classes of functions with a special property, rather than a property that a given function may or may not have. Furthermore, even within the study of sequences of functions, the condition given by Eq.~\eqref{Eq: condition superoscillating sequence} is somewhat restrictive\footnote{Although this definition of superoscillating sequence can be generalized~\cite{Colombo2018,Colombo2019} to allow convergence to non-monochromatic functions, it is still limited to sequences of functions. Additionally, the generalization seems to be less practical: the (monochromatic-convergent) version presented here is by far the most widely used in the literature, even by the same authors of~\cite{Colombo2018,Colombo2019}.}, since it does not allow the sequence to converge to any oscillating function, but specifically a monochromatic one. This, as we will see, excludes from the definition some functions that are arguably superoscillating from the qualitative point of view.


On the other hand, a popular attempt to characterize superoscillations for a generic function that relies on the qualitative notion of superoscillations was first considered in~\cite{Berry1995}, and has been further used either as a definition or as a measure in later references, such as, e.g.,~\cite{Berry2006,Dennis2008,Berry2008,Dennis2009,Berry2016}: in these references, a complex function $u$ is defined to be superoscillating if its \textit{local gradient}, or \textit{local wavenumber}, 
\begin{equation}\label{Eq: local wavenumber}
k(x) = \partial_{x} \arg{u(x)} = \Im \frac{\partial_{x} u(x)}{u(x)},
\end{equation}
is higher than the frequency of its fastest Fourier component. $|k(x)|$ is then suggested as a measure of the degree of superoscillatory behaviour: the larger $|k(x)|$, the more significant the superoscillation. This definition has the clear advantage of its simplicity, and that it may, at first sight, seem to capture the intuition of what superoscillations ought to look like.

However, this definition has three critical issues. The first issue is that the local wavenumber defined in Eq.~\eqref{Eq: local wavenumber} is identically zero when evaluated for real-valued functions, regardless of their behaviour. Yet, we should be able to talk about a real-valued function being superoscillating, just in the same way we say that the real part of $g(x,a,N)$ is superoscillating (cf. Fig.~\ref{Fig: Classic superoscillation example}). For physical superoscillations in, e.g., optics, the definition has been used by expressing the oscillating function as the real part of a complex-valued function, and then applying this definition to the latter (see, e.g.,~\cite{Dennis2008}). However, this procedure is ambiguous, since a real-valued function can have more than one complex-valued function of which it is the real part, and each of these complex functions can yield a different value of the local wavenumber: consider, for example~\cite{Chojnacki2016},
\begin{equation}\label{Eq: example product of cosines}
h(x,m,n) = \cos(mx) \cos(nx),
\end{equation}
for $m , n \in \mathbb{Z}$. Then, we have that
\begin{equation}
u_1(x,m,n) = \frac{e^{\ii(m+n)x} + e^{\ii(m-n)x}}{2}
\end{equation}
and
\begin{equation}\label{Eq: result for h(x,1,2) with Hilbert transform}
u_2(x,m,n) = \frac{e^{\ii(m+n)x} + e^{-\ii(m-n)x}}{2}
\end{equation}
satisfy that
\begin{equation}
h(x,m,n) = \Re\{u_1(x,m,n)\}=\Re\{u_2(x,m,n)\}.
\end{equation}
However, for $n \neq m$, 
\begin{equation}\label{Eq: discrepancy local wavenumbers}
k_1(x) = \Im \frac{\partial_x u_1}{u_1} = m \neq n = \Im \frac{\partial_x u_2}{u_2} = k_2(x).
\end{equation}
This illustrates the ambiguity of this way of defining the local wavenumber for real-valued functions (and therefore of this definition of superoscillatory behaviour), also casting doubts on the ability to use $|k(x)|$ as a measure of the degree of superoscillation. 

\begin{figure}
\includegraphics[scale=0.9]{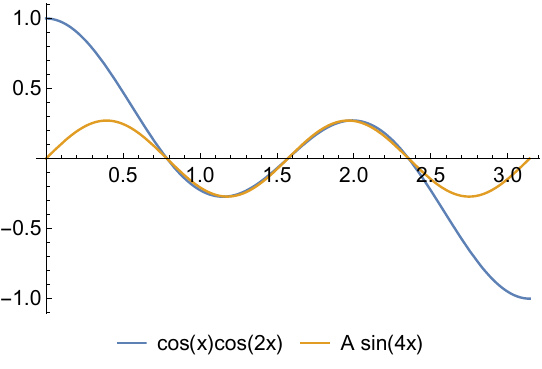}
\vspace{-0.2cm}
\caption{Comparison between $h(x, 1, 2)$ and a function proportional to $\sin(4x)$ in the interval $[0,\pi]$. The plot suggests that $h(x,1,2)$ is superoscillating around $x=\pi/2$.}
\label{Fig: Cosines superoscillation example}
\vspace{-0.2cm}
\end{figure}

The previous issue could still be addressed by fixing a specific method to construct a complex-valued function from the real signal: for instance, it is standard in signal theory to obtain the local phase of a real function using the Hilbert transform~\cite{King2009HTch18}. Specifically, given an analytic real function $v(x)$, we can prescribe the complex-valued function $u$ associated with $v$ to be
\begin{equation}\label{Eq: Hilbert transform complex function}
u(x) = v(x) + \ii \text{H}[v](x), 
\end{equation}
where $\text{H}[v]$ denotes the Hilbert transform\footnote{The Hilbert transform of a function $v$ is defined as~\cite{King2009HTch3}
\begin{equation}
\text{H}[v](x) = \frac{1}{\pi} \text{PV} \!\int_{-\infty}^\infty\!\! \diff t\, \frac{v(t)}{x-t} = \frac{1}{\pi}\!\!\! \lim_{\phantom{a}\epsilon\to0^+} \int_{|x-t|>\epsilon} \!\!\diff t\,\frac{v(t)}{x-t},
\end{equation}
and its Fourier transform satisfies~\cite{King2009HTch5}
\begin{equation}
\mathcal{F}\{\text{H}[v]\}(\omega) = - \ii \sgn\omega \, \mathcal{F}\{v\}(\omega).
\end{equation}
} of $v$. This particular choice has indeed been used in the context of superoscillations (see, e.g.,~\cite{McCaul2022}). However, even if we use this method, there is a second issue that this definition does not address: it still does not capture some superoscillatory behaviours. To illustrate this, consider once more the function in Eq.~\eqref{Eq: example product of cosines}, in the particular case when $m=1$ and $n=2$. From Eq.~\eqref{Eq: Hilbert transform complex function}, we find that the complex-valued function associated with $h(x,1,2)$ according to the Hilbert transform method is the one given precisely by Eq.~\eqref{Eq: result for h(x,1,2) with Hilbert transform}, i.e., $u_2(x,1,2)$. Eq.~\eqref{Eq: discrepancy local wavenumbers} then yields a constant local wavenumber, $k(x)=2$, while $h(x,1,2)$ is easily seen to be bandlimited to the interval $[-3,3]$. This means that the definition of superoscillations in terms of the local wavenumber would not classify it as a superoscillating function. However, in Fig.~\ref{Fig: Cosines superoscillation example} we can see that $h(x,1,2)$ oscillates almost synchronically with a function proportional\footnote{The proportionality constant is simply picked to match the amplitude of $h(x,1,2)$ around the superoscillating interval.} to $\sin(4x)$, whose frequency is outside of the Fourier spectrum of $h(x,1,2)$, around $x=\pi/2$. This leads to the conclusion that $h(x,1,2)$ is indeed a superoscillating function, and yet its superoscillatory behaviour is not captured by the local wavenumber. To reinforce this point, consider also the square of a shifted cosine, which was given as an example of superoscillating function (around $x=0$) in~\cite{Remez2015,Berry2019Roadmap,Rogers2020}:
\begin{equation}\label{Eq: shifted square cosine}
h_s(x,m) = \big[ \cos(mx) - s \big]^2,
\end{equation}
for some $0<s<1$. We can write
\begin{equation}
h_s(x,m) = \Re\bigg\{  \frac{1}{2} e^{2 \ii m x} - 2 s e^{\ii m x} + \frac{1}{2} + s^2 \bigg\},
\end{equation}
where the complex-valued function inside the real part was obtained using the Hilbert transform method in Eq.~\eqref{Eq: Hilbert transform complex function}. Using this prescription to evaluate the local wavenumber, it can be shown that 
\begin{equation}
k(x) \leq m,
\end{equation}
for all $x \in \R{}$. This means again that with this definition of superoscillations we could end up classifying this widely-accepted superoscillating function as non-superoscillating.

The third problem of the definition of superoscillations based on the local wavenumber is precisely its local nature, which implies that it can generally apply to superoscillating intervals that are ``too short'', in the sense that the function oscillates much less than half a cycle. For instance, it was shown in~\cite{Berry2008} that one-dimensional monochromatic waves can exhibit superoscillations according to this definition, but these happen very briefly in the neighbourhood of points where the function is close to zero. Arguably, we should not expect to be able to use this kind of extremely brief superoscillation in practical applications, such as, e.g., encoding information~\cite{Kempf2000Beethoven,Ferreira2006}, or superresolution~\cite{Zheludev2008,Rogers2013Imaging}. Intuitively, the reason for this is that, even though it is not impossible that some of these functions are used to, e.g., encode \textit{some} messages---in the same way a string of zeros can encode \textit{one} message---, it seems reasonable to expect that to be able to encode an alphabet in a wave with a certain rate, one needs at least half-wave oscillations within the distance between one sample and the next one, otherwise an encoding scheme that uses superoscillations can hardly take advantage of them to increase the information density. This is even more so for optical applications of superoscillations such as superresolution, where overcoming the diffraction limit relies on having shorter but full wavelengths for some period of time. Because of this, we argue that those functions displaying a superoscillatory behaviour that is not long-lived enough should not be considered superoscillating functions for practical purposes.

This analysis reveals that there is room to propose a new definition that can be universally applied to any given function. This is the problem that we address in the following section.

\section{Identifying superoscillations}\label{Section: Identifying 1D superoscillations}

In this section, we introduce a formal definition of superoscillatory behaviour that does not exhibit the problems of previous proposals. First, we will introduce the two criteria on which the definition relies. As we will see, these two criteria allow us to discriminate whether a given real-valued one-dimensional function is superoscillating or not in certain intervals. The definition then generalizes this to arbitrary intervals.

The first criterion is designed to capture superoscillations around a specific horizontal reference line, $y=c$.

\begin{criterion}[Sine criterion for $y=c$]\label{Sine criterion}
Given a real-valued one-dimensional function $f$ which is bandlimited to \mbox{$[-\pi W, \pi W] \subset \R{}$}, and given \mbox{$b_1,b_2 \in \R{}$} such that \mbox{$b_1<b_2$} and \mbox{$f(b_1)=f(b_2)=c$}, let
\begin{equation}\label{Eq: sine coefficients}
a_k = \int_{b_1}^{b_2} \diff x \, \big[ f(x) - c \big] \sin\bigg( \frac{\pi k (x - b_1)}{b_2-b_1} \bigg),
\end{equation}
for each $k\in\mathbb{N}$, be the Fourier sine coefficients over the interval $[b_1,b_2]$ around the horizontal line $y=c$. If we denote with $k_0 \in \mathbb{N}$ the smallest natural number that satisfies $k_0 >  W (b_2-b_1)$, we define
\begin{equation}\label{Eq: Qs definition}
Q_{\textnormal{sin}} \coloneqq \sqrt{\frac{\sum_{k=k_0}^{\infty}|a_k|^2}{\sum_{k=1}^\infty|a_k|^2}}.
\end{equation}
Then, we say that $f$ is superoscillating around $y=c$ in the interval $[b_1,b_2]$ if $Q_\textnormal{sin}>1/2$.
\end{criterion}

Notice that the Fourier sine coefficients evaluated according to Eq.~\eqref{Eq: sine coefficients} are exactly the Fourier coefficients of a periodic function with period (at most) $2(b_2-b_1)$ that is odd with respect to the vertical line $x=b_1$. In a way, we are evaluating a ``local Fourier spectrum''\footnote{By \textit{local Fourier spectrum} we mean that this is the Fourier spectrum of the periodic (odd with respect to $x=b_1$) extension of the function $f$ in the interval $[b_1,b_2]$, as opposed to the (in general continuous) Fourier spectrum of the whole function $f$ in $\R{}$.} of the function with the information available in the interval $[b_1,b_2]$. Thus, the quantity $Q_\text{sin}$ is simply the ratio between two quantities: i) the (2-norm\footnote{The choice of the 2-norm to measure the `weight' of each contribution is based on the fact that it is invariant under Fourier transform. This is also the reason why we use $1/2$ as the threshold for superoscillatory behaviour, i.e., when more than half the weight of the Fourier spectrum is outside the bandlimit. Nevertheless, if a particular problem demands the use of a different norm or threshold, the sine criterion (as well as the upcoming cosine criterion) can be trivially modified to adapt to these different choices.}) weight of the local Fourier sine coefficients associated with frequencies outside the total Fourier spectrum of the function $f$, and ii) the weight of all the local Fourier sine coefficients together. The criterion then establishes that if the superoscillating contributions (i.e., the ones with a frequency beyond the bandlimit) stand for more than half the weight of the contributions of all the Fourier sine coefficients, then the function should be considered to display a superoscillatory behaviour around $y=c$ in the interval $[b_1,b_2]$. 

It is worth remarking that this criterion cannot be applied to just any interval, but specifically, those for which the function has value $c$ at its endpoints. Thus, the sine criterion alone might not be enough to characterize superoscillations when there is not a specific horizontal line of reference around which the function oscillates. That is why we introduce a second criterion, designed to capture superoscillations precisely in such scenarios.

\begin{criterion}[Cosine criterion]\label{Cosine criterion}
Given a real-valued one-dimensional function $f$ which is bandlimited to $[-\pi W, \pi W] \subset \R{}$, and given \mbox{$b_1,b_2 \in \R{}$} such that \mbox{$b_1<b_2$} and \mbox{$f'(b_1)=f'(b_2)=0$}, let
\begin{equation}\label{Eq: cosine coefficients}
a_k = \int_{b_1}^{b_2} \diff x \, f(x) \cos\bigg( \frac{\pi k (x-b_1)}{b_2-b_1} \bigg),
\end{equation}
for each $k\in\mathbb{N}$, be the Fourier cosine coefficients over the interval $[b_1,b_2]$. If we denote with $k_0 \in \mathbb{N}$ the smallest natural number that satisfies $k_0 >  W (b_2-b_1)$, we define
\begin{equation}\label{Eq: Qc definition}
Q_\textnormal{cos} \coloneqq \sqrt{\frac{\sum_{k=k_0}^{\infty}|a_k|^2}{\sum_{k=1}^\infty|a_k|^2}}.
\end{equation}
Then, we say that $f$ is superoscillating in the interval $[b_1,b_2]$ if $Q_\textnormal{cos}>1/2$.
\end{criterion}

Here, the Fourier cosine coefficients evaluated according to Eq.~\eqref{Eq: cosine coefficients} are exactly the Fourier coefficients of a periodic function with period (at most) $2(b_2-b_1)$ that is even with respect to the vertical line $x=b_1$. In a way, we are again evaluating a ``local Fourier spectrum''\footnote{Now considering the periodic (even with respect to $x=b_1$) extension of the function $f$ in the interval $[b_1,b_2]$.} of the function with the information available in the interval $[b_1,b_2]$. The quantity $Q_\text{cos}$ is, in a completely analogous fashion, a ratio between the weight of the Fourier coefficients associated with superoscillating modes and the total weight of all the coefficients together except for the zero mode (which accounts for the possible vertical translation of the function, without altering its spectrum elsewhere). Then, we consider the function to be superoscillating in the interval if the superoscillating modes have more overall weight in the local Fourier decomposition than the ones within the bandlimited Fourier spectrum of the function. This criterion, which can only be applied to intervals whose endpoints are local extrema of the function, seems intuitively less restrictive than the sine criterion. The sine criterion is indeed tailored for those situations in which the \textit{only} oscillations of interest are about a particular horizontal line, while the cosine criterion allows for oscillations that do not necessarily have a prespecified horizontal line they oscillate about. Nevertheless, the cosine and the sine criteria are generally applied to different intervals, and therefore it is not possible in principle to prove an inequality that establishes a hierarchy between them.

Once we have introduced the sine and cosine criteria, we can give the following definitions of superoscillating function.
\begin{definition*}[Superoscillating function in an interval]
A real-valued one-dimensional bandlimited function $f$ is considered to display superoscillations around $y=c$ in an interval $I$ if there exist intervals $I_1$ and $I_2$ such that \mbox{$I_1 \subseteq I \subseteq I_2$}, and the sine criterion for $y=c$ is satisfied for both $I_1$ and $I_2$. Alternatively, $f$ is considered to display superoscillations in $I$ if there exist intervals $I_1$ and $I_2$ such that \mbox{$I_1 \subseteq I \subseteq I_2$}, and the cosine criterion is satisfied for both $I_1$ and $I_2$.
\end{definition*}
As a remark, note that the inclusions $I_1 \subseteq I \subseteq I_2$ are not necessarily strict, and therefore if $I$ is an interval that satisfies one of the two criteria, then the definition is automatically satisfied by choosing $I_1 = I_2 = I$. 

This definition addresses at first glance some of the issues of the traditional criteria, since 1) it is well defined for any given real-valued one-dimensional bandlimited function, and 2) they can only answer affirmatively the question of whether a function is superoscillating or not in intervals that are long enough to capture at least half a (super)oscillation. This property rules out by fiat those functions that display superoscillations that are too short-lived to be of practical use. Moreover, in the next section we will see that this definition identifies superoscillatory behaviours that the other definitions do not capture. We will also see that, since $Q_\text{sin}$ and $Q_\text{cos}$ represent overall contributions of the superoscillating modes (and in particular they are exactly zero for sines and cosines), these values can be used as measures of the degree of superoscillatory behaviour displayed by the function of interest. 

Notice that this definition is intended to capture superoscillations about horizontal lines (whether specified, as in the sine criterion, or unspecified, as in the cosine one). However, it is possible to readily generalize the definition to account for superoscillations about sloped lines, or even around an arbitrary polynomial $p(x)$ by virtue of applying the criteria on the function $\tilde f(x)=f(x)-p(x)$.



\vspace{-0.2cm}

\section{Examples}\label{Section: Examples}

In this section, we apply the definition of superoscillating function given in Sec.~\ref{Section: Identifying 1D superoscillations} to a variety of examples, both superoscillating and non-superoscillating, to check how it correctly captures the qualitative description of whether a function has superoscillations or not in a particular interval. For greater clarity, we will discuss the sine and cosine criteria separately. 

For the sake of simplicity, here we only consider examples in which the oscillation happens about $y=0$. Also for convenience, we will analyze functions that are odd using the sine criterion and those that are even using the cosine criterion.

\vspace{-0.3cm}

\subsection{Sine criterion}\label{Subsection: Sine criterion}

\begin{figure*}
\begin{center}
\includegraphics[scale=0.87]{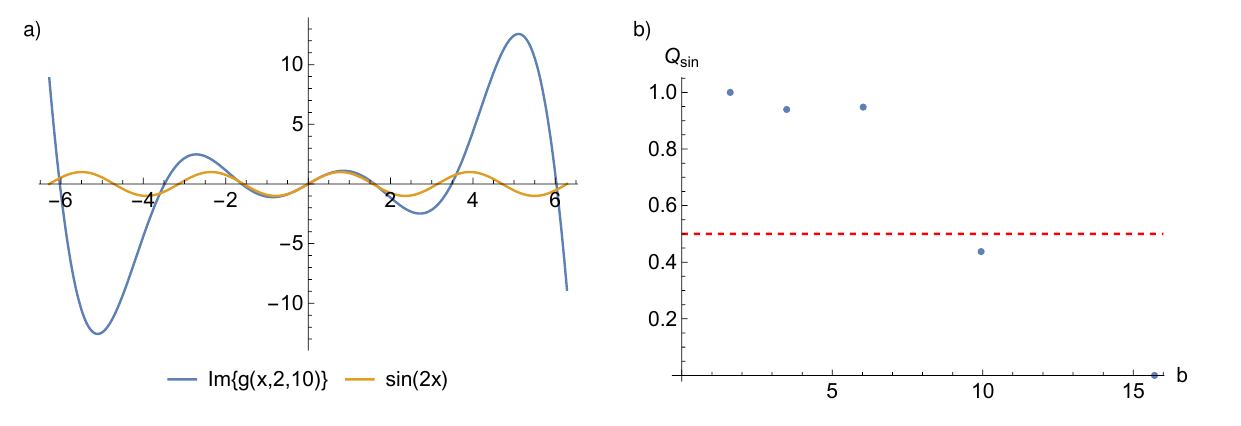}
\vspace{-0.8cm}
\caption{a) Comparison between the imaginary part of $g(x,2,10)$ and $\sin(2x)$ in the interval $[-2\pi,2\pi]$, showing that $\Im\{g(x,2,10)\}$ is superoscillating around $x=0$. b) Values of $Q_\text{sin}$ for intervals of the form $[0,b]$, as a function of the right endpoint $b>0$. The value of $Q_\text{sin}$ was plotted for consecutive zeros, until two points were reached for which $Q_\text{sin}<1/2$, indicating the end of the superoscillating interval.}
\label{Fig: Img10}
\end{center}
\vspace{-0.5cm}
\end{figure*}

\begin{figure*}
\begin{center}
\includegraphics[scale=0.87]{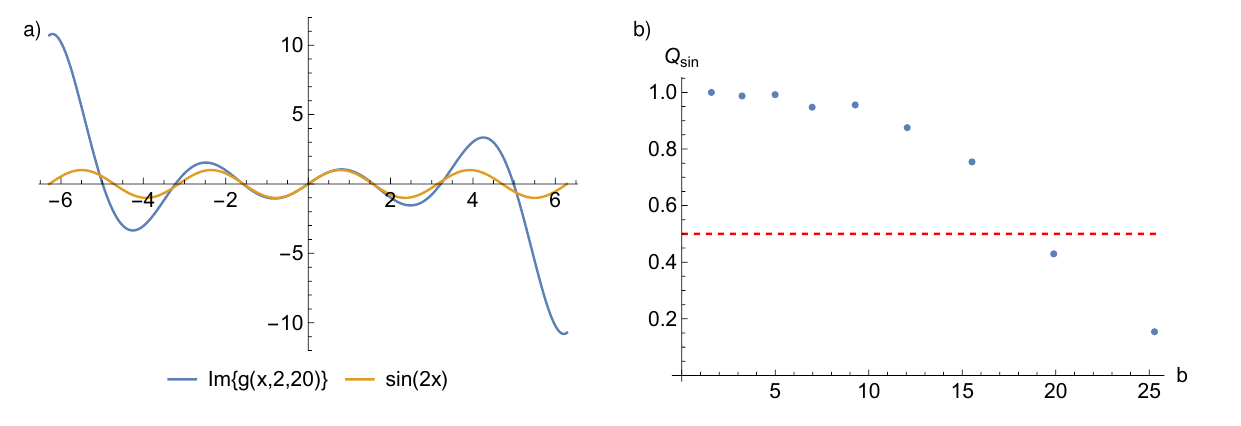}
\vspace{-0.8cm}
\caption{a) Comparison between the imaginary part of $g(x,2,20)$ and $\sin(2x)$ in the interval $[-2\pi,2\pi]$, showing that $\Im\{g(x,2,20)\}$ is superoscillating around $x=0$. b) Values of $Q_\text{sin}$ for intervals of the form $[0,b]$, as a function of the right endpoint $b>0$. The value of $Q_\text{sin}$ was plotted for consecutive zeros, until two points were reached for which $Q_\text{sin}<1/2$, indicating the end of the superoscillating interval.}
\label{Fig: Img20}
\end{center}
\vspace{-0.5cm}
\end{figure*}

The first example that we consider here is the imaginary part of $g(x,a,N)$, as defined in Eq.~\eqref{Eq: definition classic example}. In particular, we will examine the cases $N=10$ and $N=20$, for $a=2$, which are represented in Figs.~\ref{Fig: Img10}a and~\ref{Fig: Img20}a. These functions are bandlimited to $[-1,1]$, but they are known to display superoscillations around $x=0$, as the figures illustrate. For these functions, we evaluated $Q_\text{sin}$ in intervals of the form $[0,b]$, for several consecutive zeros $b>0$ as right endpoints. Notice that since these functions are odd, the value of $Q_\text{sin}$ in the intervals $[0,b]$ and $[-b,b]$ is the same. The results for $Q_\text{sin}$ are plotted in Figs.~\ref{Fig: Img10}b and~\ref{Fig: Img20}b, and support that the functions are superoscillating around $x=0$. Notice that even though the deviation of these functions from the monochromatic $\sin(2x)$ is evident relatively soon (after the first or second zero), the value of $Q_\text{sin}$ witnesses superoscillatory behaviour for much longer than this. The reason is that even if these functions stop resembling a monochromatic function relatively close to the origin, they superoscillate (i.e., locally oscillate faster than the fastest Fourier component) for longer than this. This behaviour is captured as superoscillating by the definition we propose here, but is missed by the definition based on superoscillating sequences, which estimates the superoscillating behaviour to happen only where the function is approximately monochromatic (i.e., the region which is called in~\cite{Berry2006} of `fast superoscillations', corresponding here to $|x|<\sqrt{N}$).

\begin{figure*}
\begin{center}
\includegraphics[scale=0.87]{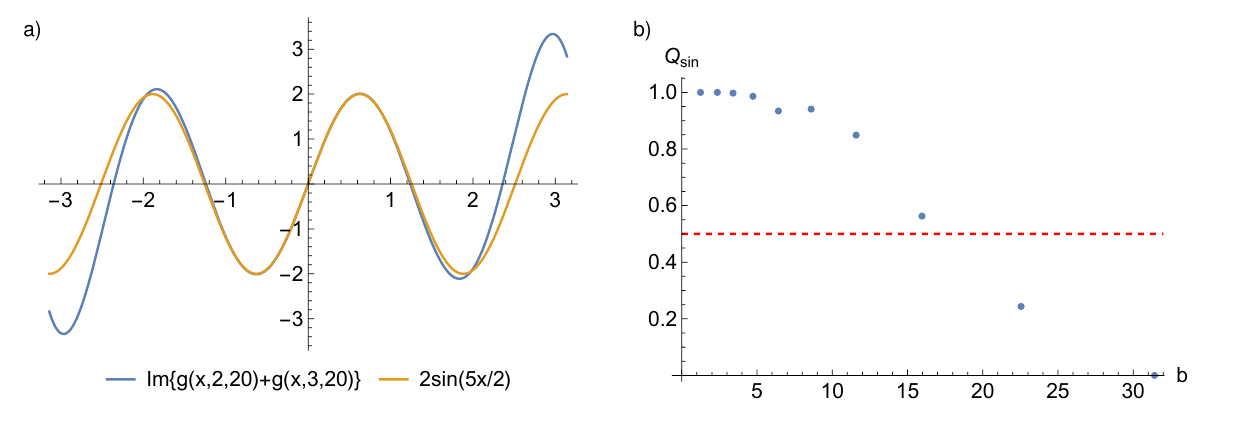}
\vspace{-0.8cm}
\caption{a) Comparison between the imaginary part of $g(x,2,20)+g(x,3,20)$ and $2\sin(5x/2)$ in the interval $[-\pi,\pi]$, showing that $\Im\{g(x,2,20)+g(x,3,20)\}$ is superoscillating around $x=0$. b) Values of $Q_\text{sin}$ for intervals of the form $[0,b]$, as a function of the right endpoint $b>0$. The value of $Q_\text{sin}$ was plotted for consecutive zeros, until two points were reached for which $Q_\text{sin}<1/2$, indicating the end of the superoscillating interval.}
\label{Fig: Sumg}
\end{center}
\vspace{-0.5cm}
\end{figure*}

An even better example of how the definition based on superoscillating sequences could not capture general enough superoscillatory behaviours is the family of functions given by the sum \mbox{$g(x,2,N)+g(x,3,N)$}, which is bandlimited to $[-1,1]$, and whose imaginary part for the case $N=20$ is represented in Fig.~\ref{Fig: Sumg}a. Clearly, 
\begin{equation}
\lim_{N\to\infty} g(x,2,N) + g(x,3,N) = e^{2\ii x} + e^{3 \ii x},
\end{equation}
which is not a monochromatic function. Yet, Fig.~\ref{Fig: Sumg}a shows that \mbox{$\Im\{g(x,2,N)+g(x,3,N)\}$} oscillates synchronically with $2\sin(5 x/2)$ around $x=0$, and this superoscillatory behaviour is corroborated by the values of $Q_\text{sin}$ plotted in Fig.~\ref{Fig: Sumg}b. Notice that, near $x=0$,
\begin{equation}
\Im\{g(x,2,N) + g(x,3,N)\} \approx 2 \sin \frac{5x}{2} \cos \frac{x}{2}. 
\end{equation}
Thus, close enough to the origin the cosine factor is approximately 1, and the function can be approximated just by the sine. This explains the synchronic oscillation of the function and $2\sin(5x/2)$, but also makes it clear that, even if it is small, the cosine factor prevents the sequence \mbox{$g(x,2,N)+g(x,3,N)$} from converging to $2\sin(5x/2)$ in any compact interval containing $x=0$, even though it resembles to it.

\begin{figure*}
\begin{center}
\includegraphics[scale=0.87]{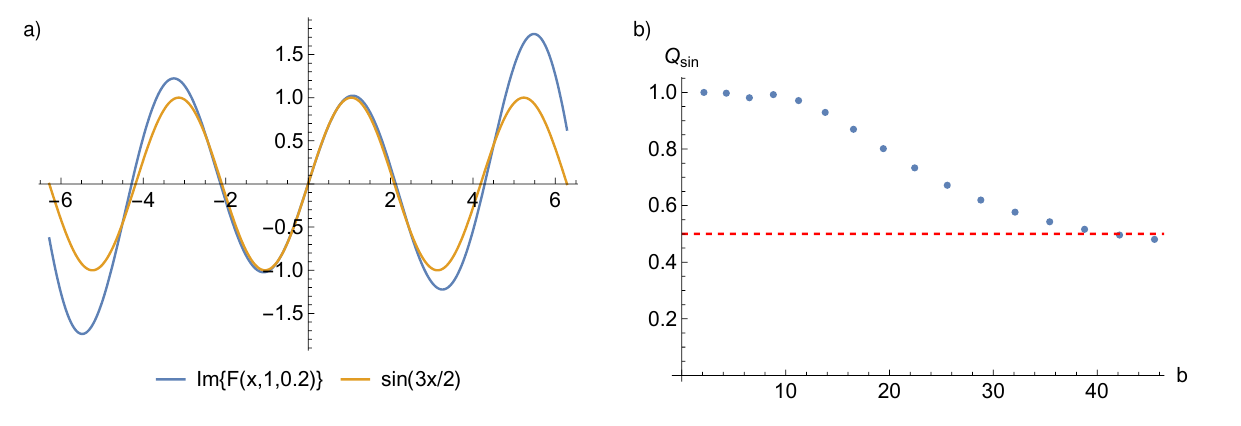}
\vspace{-0.8cm}
\caption{a) Comparison between the imaginary part of $F(x,1,0.2)$ and $\sin(3x/2)$ in the interval $[-2\pi,2\pi]$, showing that $\Im\{F(x,1,0.2)\}$ is superoscillating around $x=0$. b) Values of $Q_\text{sin}$ for intervals of the form $[0,b]$, as a function of the right endpoint $b>0$. The value of $Q_\text{sin}$ was plotted for consecutive zeros, until two points were reached for which $Q_\text{sin}<1/2$, indicating the end of the superoscillating interval.}
\label{Fig: ImF}
\end{center}
\vspace{-0.5cm}
\end{figure*}

The next example we consider is $\Im\{F(x,1,0.2)\}$, where,
\begin{align}
F(x,A,\delta) = & \frac{\erf\Big(\frac{\ii A+2 \ii \delta ^2 x+2}{\delta  \sqrt{2+2 \ii \delta ^2 x}}\Big)+\erf\Big(\frac{-\ii A+2 \ii \delta ^2 x+2}{\delta  \sqrt{2+2 \ii \delta ^2 x}}\Big) }{2 \sqrt{1+\ii \delta ^2 x}}\nonumber \\
& \times \exp \left(\frac{\ii x \left(A^2+2 \ii \delta ^2 x+2\right)}{2 \left(1+\ii \delta ^2 x\right)}\right).\label{Eq: Fexample}
\end{align}
$F$ is shown to be bandlimited in $[-1,1]$ in~\cite{Berry1995}, and it is argued to superoscillate around $x=0$, as its graph in Fig.~\ref{Fig: ImF}a illustrates. The evaluation of $Q_\text{sin}$ yields the results plotted in Fig.~\ref{Fig: ImF}b, which again successfully characterizes the superoscillatory behaviour.

\begin{figure*}
\begin{center}
\includegraphics[scale=0.87]{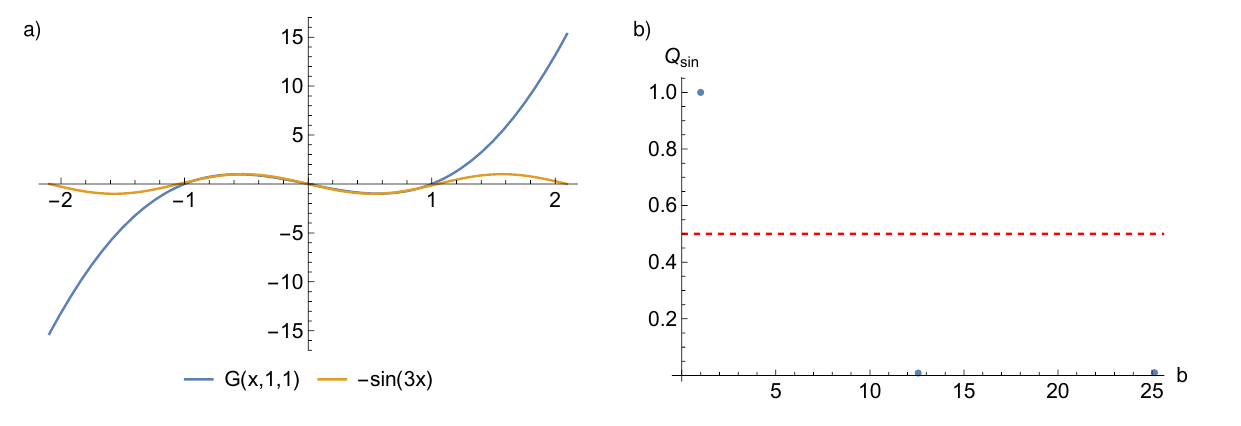}
\vspace{-0.8cm}
\caption{a) Comparison between $G(x,1,1)$ and $-\sin(3x)$ in the interval $[-2\pi/3,2\pi/3]$, showing that $G(x,1,1)$ is superoscillating around $x=0$. b) Values of $Q_\text{sin}$ for intervals of the form $[0,b]$, as a function of the right endpoint $b>0$. The value of $Q_\text{sin}$ was plotted for consecutive zeros, until two points were reached for which $Q_\text{sin}<1/2$, indicating the end of the superoscillating interval.}
\label{Fig: G}
\end{center}
\vspace{-0.5cm}
\end{figure*}

One more example we use to test the criterion is
\begin{equation}
G(x, s, D) = \frac{3\sqrt{3}}{2} \left( \frac{x^3}{s^3} -\frac{x}{s} \right) \operatorname{sinc}^4(t/4D).
\end{equation}
The particular case $G(x,1,1)$ is represented in Fig.~\ref{Fig: G}a. In~\cite{Chremmos2015}, $G$ was shown to be bandlimited to $[-1,1]$, and argued to be superoscillating around $x=0$, for $D\geq1$ and $s\leq 1$. This superoscillatory behaviour is indeed witnessed by the $Q_\text{sin}$ values plotted in Fig.~\ref{Fig: G}b. 

\begin{figure}
\includegraphics[scale=0.87]{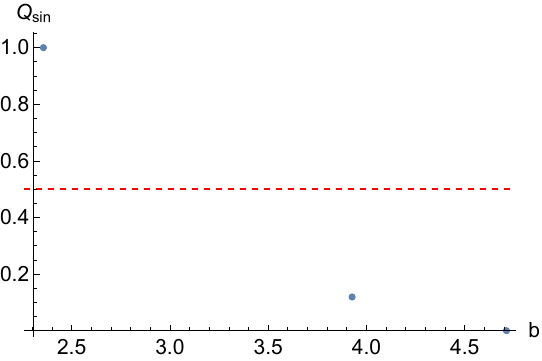}
\caption{Values of $Q_\text{sin}$ for $h(x,1,2)$ in intervals of the form $[\pi/2,b]$, as a function of the right endpoint $b>\pi/2$. The value of $Q_\text{sin}$ was plotted for consecutive zeros, until two points were reached for which $Q_\text{sin}<1/2$, indicating the end of the superoscillating interval.}
\label{Fig: Qh}
\vspace{-0.3cm}
\end{figure}

Finally, we can revisit $h(x,1,2)$, which was given in Eq.~\eqref{Eq: example product of cosines} and represented in Fig.~\ref{Fig: Cosines superoscillation example}, and which we showed in Sec.~\ref{Section: The challenge of defining superoscillations} that the local wavenumber definition fails to identify as superoscillating. Looking at its graph, the function clearly displays a superoscillatory behaviour around $x=\pi/2$. Since the function is odd with respect to this point, we evaluate $Q_{\text{sin}}$ for intervals $[\pi/2,b]$, for consecutive zeros of the function $b>\pi/2$. The results are plotted in Fig.~\ref{Fig: Qh}, showing that the sine criterion does identify the superoscillatory behaviour of this function around $x=\pi/2$. 

\vspace{-0.2cm}

\subsection{Cosine criterion}\label{Subsection: Cosine criterion}

\begin{figure*}
\begin{center}
\includegraphics[scale=0.87]{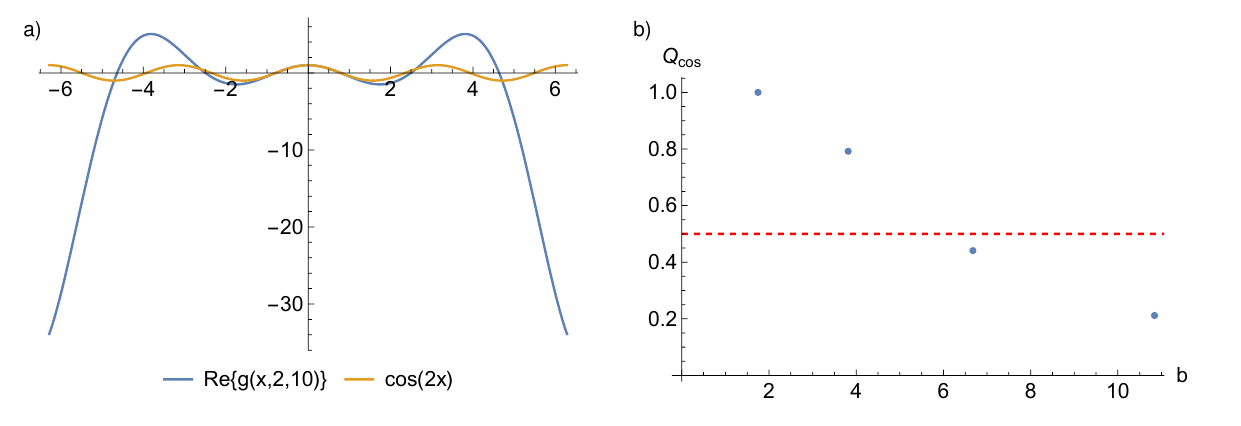}
\vspace{-0.8cm}
\caption{a) Comparison between the real part of $g(x,2,10)$ and $\cos(2x)$ in the interval $[-2\pi,2\pi]$, showing that $\Re\{g(x,2,10)\}$ is superoscillating around $x=0$. b) Values of $Q_\text{cos}$ for intervals of the form $[0,b]$, as a function of the right endpoint $b>0$. The value of $Q_\text{cos}$ was plotted for consecutive extremum points, until two points were reached for which $Q_\text{cos}<1/2$, indicating the end of the superoscillating interval.}
\label{Fig: Reg10}
\end{center}
\vspace{-0.5cm}
\end{figure*}

\begin{figure*}
\begin{center}
\includegraphics[scale=0.87]{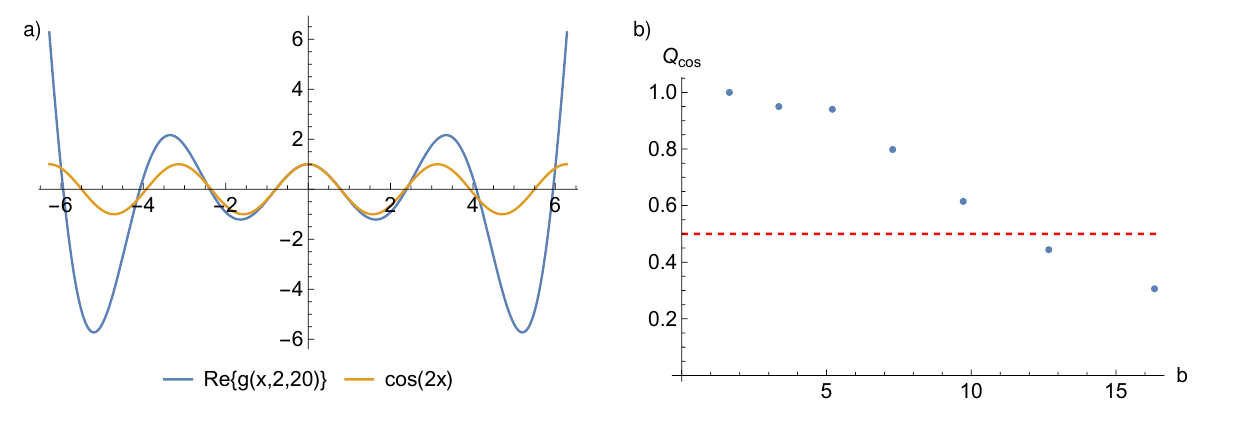}
\vspace{-0.8cm}
\caption{a) Comparison between the real part of $g(x,2,20)$ and $\cos(2x)$ in the interval $[-2\pi,2\pi]$, showing that $\Re\{g(x,2,20)\}$ is superoscillating around $x=0$. b) Values of $Q_\text{cos}$ for intervals of the form $[0,b]$, as a function of the right endpoint $b>0$. The value of $Q_\text{cos}$ was plotted for consecutive extremum points, until two points were reached for which $Q_\text{cos}<1/2$, indicating the end of the superoscillating interval.}
\label{Fig: Reg20}
\end{center}
\vspace{-0.5cm}
\end{figure*}

The first example that we consider here is $\Re\{g(x,a,N)\}$. In particular, we study the cases $N=10$ and $N=20$, for $a=2$, i.e., the real counterparts of the functions we examined in Sec.~\ref{Subsection: Sine criterion},  which are represented in Figs.~\ref{Fig: Reg10}a and~\ref{Fig: Reg20}a, respectively. For these functions, we evaluated $Q_\text{cos}$ in intervals of the form $[0,b]$, for several consecutive extremum points $b>0$ as right endpoints. Notice that, since these functions are even, the value of $Q_\text{cos}$ for the intervals $[0,b]$ and $[-b,b]$ are the same. The values of $Q_\text{cos}$ are plotted in Figs.~\ref{Fig: Reg10}b and~\ref{Fig: Reg20}b, showing that the cosine criterion correctly identifies the superoscillatory behaviour around $x=0$.

\begin{figure*}
\begin{center}
\includegraphics[scale=0.87]{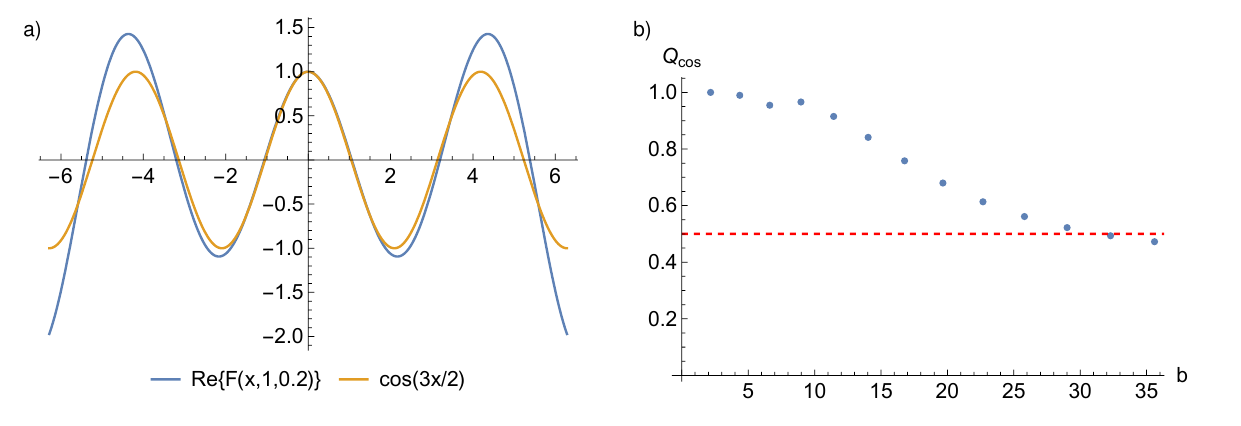}
\vspace{-0.8cm}
\caption{a) Comparison between the real part of $F(x,1,0.2)$ and $\cos(3x/2)$ in the interval $[-2\pi,2\pi]$, showing that $\Re\{F(x,1,0.2)\}$ is superoscillating around $x=0$. b) Values of $Q_\text{cos}$ for intervals of the form $[0,b]$, as a function of the right endpoint $b>0$. The value of $Q_\text{cos}$ was plotted for consecutive extremum points, until two points were reached for which $Q_\text{cos}<1/2$, indicating the end of the superoscillating interval.}
\label{Fig: ReF}
\end{center}
\vspace{-0.5cm}
\end{figure*}

The next example we consider here is $\Re\{F(x,1,0.2)\}$, once more the real part of a function whose imaginary part we analyzed in Sec.~\ref{Subsection: Sine criterion}, with $F(x,A,\delta)$ being defined in Eq.~\eqref{Eq: Fexample}. Like its imaginary counterpart, this function is bandlimited to $[-1,1]$ and superoscillating around $x=0$, as its graph in Fig.~\ref{Fig: ReF}a illustrates. The values of $Q_\text{cos}$, which are plotted in Fig.~\ref{Fig: ReF}b, indeed report this superoscillatory behaviour.

\begin{figure*}
\begin{center}
\includegraphics[scale=0.87]{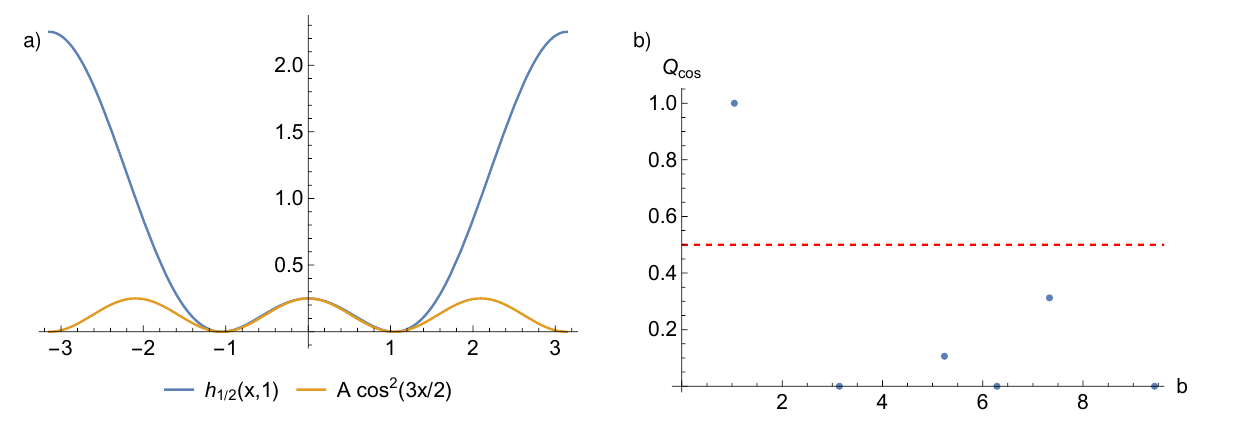}
\vspace{-0.8cm}
\caption{a) Comparison between $h_{1/2}(x,1)$ and a function proportional to $\cos^2(3x/2)$ in the interval $[-\pi,\pi]$, showing that $h_{1/2}(x,1)$ is superoscillating around $x=0$. b) Values of $Q_\text{cos}$ for intervals of the form $[0,b]$, as a function of the right endpoint $b>0$. The value of $Q_\text{cos}$ was plotted for consecutive extremum points until past the centre of the next superoscillation (around $x=2\pi$), in order to show how this reflects on the value of $Q_\text{cos}$ as an increase.}
\label{Fig: shiftedsquaredcos}
\end{center}
\vspace{-0.5cm}
\end{figure*}

As our last example, we can revisit the square of the shifted cosine, which we denoted as $h_s(x,m)$ in Eq.~\eqref{Eq: shifted square cosine}. This function, represented in Fig.~\ref{Fig: shiftedsquaredcos}a for the particular case $m=1$ and $s=1/2$, was used in Sec.~\ref{Section: The challenge of defining superoscillations} as another example of superoscillation that is not captured by the definition based on the local wavenumber. The results of $Q_\text{cos}$, plotted in Fig.~\ref{Fig: shiftedsquaredcos}b, show that the definition proposed here captures the superoscillatory behaviour of the function around $x=0$. The value of $Q_\text{cos}$ drops when evaluated with the second extremum point as right endpoint. This indicates that the superoscillation is short-lived, in agreement with the graphical estimates in~\cite{Remez2015,Berry2019Roadmap}. It is also easy to observe in Fig.~\ref{Fig: shiftedsquaredcos}b that there is a noticeable increase of $Q_\text{cos}$ for the third and fifth extremum points ($b=5\pi/3$ and $7\pi/3$). This is due to the onset of a new superoscillating interval around $x = 2\pi$. While the function  $h_{1/2}(x,1)$ is not superoscillating in the whole interval $[0,5\pi/3]$, or in the whole interval $[0,7\pi/3]$, one can see how  $Q_\text{cos}$ still captures this onset. This reinforces the idea that it can be used not only as a witness, but also as a measure of the degree of superoscillatory behaviour.  

Finally, to further support the adequacy of the proposed definition of superoscillations, it is worth confronting our criteria against examples of non-superoscillating functions, to make sure they do not produce false positives. First, we can consider the case of $(\cos x)^2$, for which the value of $Q_\text{cos}$ can be easily calculated. Specifically, for intervals of the form $[0,b]$, where $b=\alpha\pi/2$ with $\alpha \in \mathbb{N}$,
\begin{align}
a_k & = \!\int_0^{\alpha\pi/2} \!\diff x (\cos x)^2 \cos\bigg( \frac{2 k x}{\alpha} \bigg) = \begin{dcases*}
\alpha \pi /4 & if $k=0$ \\
\alpha \pi / 8 & if $k=\alpha$ \\
0 & otherwise
\end{dcases*}\!\!. 
\end{align}
Since $(\cos x)^2$ is bandlimited to $[-2,2]$ (i.e., $W=2/\pi$), the smallest natural number greater than $W b = \alpha$ is $k_0=\alpha+1$, and thus $Q_\text{cos}=0$ for intervals of the form $[0,b]$, with the right endpoint $b$ being an extremum point, as the application of the cosine criterion requires. 

\begin{figure*}
\begin{center}
\includegraphics[scale=0.87]{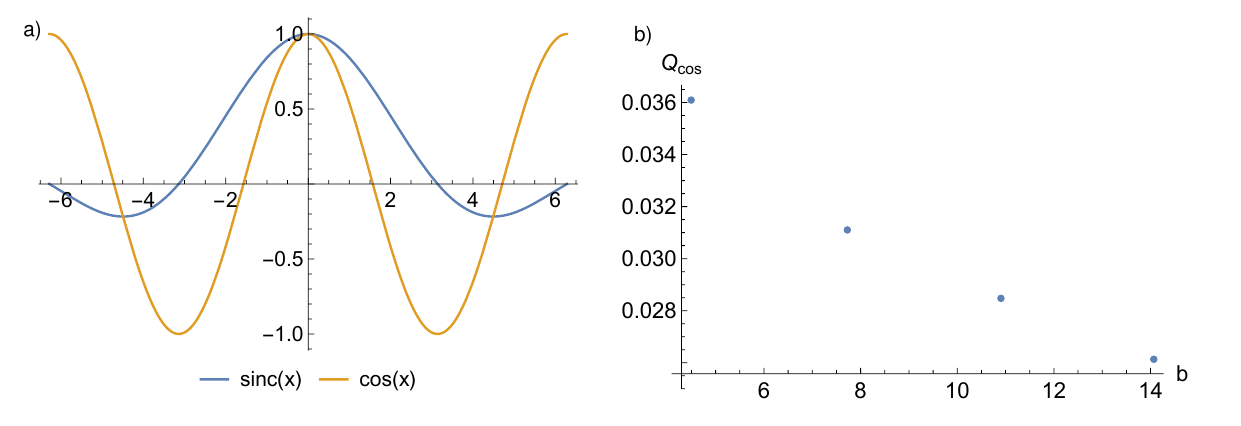}
\vspace{-0.8cm}
\caption{a) Comparison between $\operatorname{sinc}x$ and $\cos x$ in the interval $[-2\pi,2\pi]$, showing that $\operatorname{sinc} x$ is not superoscillating around $x=0$. b) Values of $Q_\text{cos}$ for intervals of the form $[0,b]$, as a function of the right endpoint $b>0$. The values of $Q_\text{cos}$ were plotted for four consecutive extremum points, showing that all of them (easily) satisfy $Q_\text{cos}<1/2$.}
\label{Fig: sinc}
\end{center}
\vspace{-0.5cm}
\end{figure*}

\begin{figure*}
\begin{center}
\includegraphics[scale=0.87]{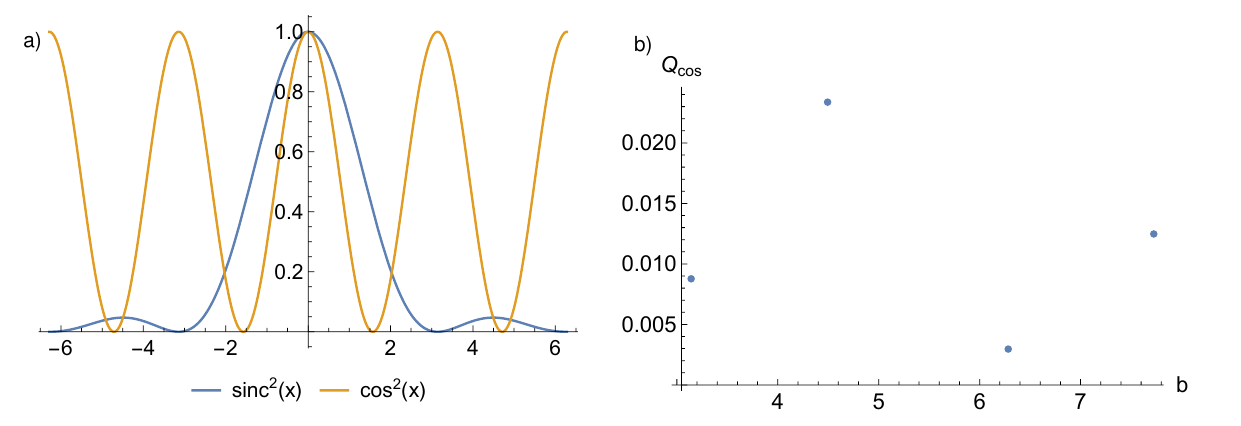}
\vspace{-0.8cm}
\caption{a) Comparison between $(\operatorname{sinc}x)^2$ and $(\cos x)^2$ in the interval $[-2\pi,2\pi]$, showing that $(\operatorname{sinc} x)^2$ is not superoscillating around $x=0$. b) Values of $Q_\text{cos}$ for intervals of the form $[0,b]$, as a function of the right endpoint $b>0$. The values of $Q_\text{cos}$ were plotted for four consecutive extremum points, showing that all of them (easily) satisfy $Q_\text{cos}<1/2$.}
\label{Fig: sinc2}
\end{center}
\vspace{-0.5cm}
\end{figure*}

We can also check that the cosine criterion does not identify any superoscillatory behaviour around $x=0$ for the functions $\operatorname{sinc}x$ and $(\operatorname{sinc}x)^2$, which are bandlimited to $[-1,1]$ and $[-2,2]$, respectively. Their graphical representations in Figs.~\ref{Fig: sinc}a and~\ref{Fig: sinc2}a show that these functions do not superoscillate, and their respective results for $Q_\text{cos}$ plotted in Figs.~\ref{Fig: sinc}b and~\ref{Fig: sinc2}b are in agreement with this qualitative judgement. 

\section{Conclusion and outlook}\label{Section: Conclusion}

\vspace{-0.1cm}

We proposed a rigorous definition of superoscillating function in an interval that also allows us to quantify the degree of such superoscillation. We illustrated through several examples that the definition captures the superoscillatory behaviour of functions that are widely accepted in the literature as superoscillating, and classifies as non-superoscillating other functions that should not be identified as such.  Moreover, it does not suffer from any of the issues that we argue are exhibited by previous proposals. 

Having a quantifier of superoscillations that is connected to their potential physical relevance can open new avenues for the formal study of superoscillations, as well as for the discovery of new superoscillating functions and methods to generate them. There is also plenty of room for generalizations, including, e.g., the prospect that the phenomenon of superoscillations can manifest itself---albeit in an approximate way---outside the realm of bandlimited functions~\cite{YuAlmostbandlimited}.


\vspace{-0.2cm}

\begin{acknowledgments}	


The authors would like to thank Achim Kempf for helpful discussions and feedback on the manuscript. YL acknowledges the support of the Undergraduate Research Award from the Institute for Quantum Computing. JPG acknowledges the support of a Mike and Ophelia Lazaridis Fellowship, as well as the support of a fellowship from ``La Caixa'' Foundation (ID 100010434, code LCF/BQ/AA20/11820043). EMM acknowledges support through the Discovery Grant Program of the Natural Sciences and Engineering Research Council of Canada (NSERC). EMM also acknowledges support of his Ontario Early Researcher award. 

\end{acknowledgments}

\onecolumngrid
\appendix


\twocolumngrid	
\bibliography{references}
	
\end{document}